%% file: ar06.tex
\documentclass{article}
\def\smallromani{\renewcommand{\theenumi}{\roman{enumi}}
\renewcommand{\labelenumi}{(\theenumi)}}

\usepackage{pstricks}  
\usepackage{amsfonts}  

\def\smallromani{\renewcommand{\theenumi}{\roman{enumi}}
\renewcommand{\labelenumi}{(\theenumi)}}

\usepackage{url}
\usepackage{amsmath}

\usepackage{color}
\usepackage{epsf}
\usepackage{eepic}
 \usepackage{epic}
\usepackage{psfig}
\usepackage{graphics}

\include{header}

\include{ecmds}

\usepackage{latexsym}
\usepackage{alltt}

\title{Stable partitions in coalitional games}

\author{Krzysztof R. Apt\footnote{CWI, Amsterdam
and
University of Amsterdam, the Netherlands} \quad
Tadeusz Radzik\footnote{Institute of Mathematics, Wroc{\l}aw University of Technology, Poland}
}

\date{}

\begin{document}

\maketitle
\begin{abstract}
  We propose a notion of a stable partition in a coalitional game that
  is parametrized by the concept of a defection function. This
  function assigns to each partition of the grand coalition a set of
  different coalition arrangements for a group of defecting players.
  The alternatives are compared using their social welfare.
  
  We characterize the stability of a partition for a number of most
  natural defection functions and investigate whether and how so
  defined stable partitions can be reached from any initial partition
  by means of simple transformations.
  
  The approach is illustrated by analyzing an example in which a set
  of stores seeks an optimal transportation arrangement.
\end{abstract}


\section{Introduction}

The problem of coalition formation has become an important research
direction in theoretical economics, notably game theory.  It has been
studied from many points of view beginning with \cite{AD74}, where the
static situation of coalitional games in the presence of a given
coalition structure (i.e., a partition)
was considered and where the issue of
coalition formation was briefly alluded to (on pages 233--234).
The early research on the subject was discussed in \cite{Gre94}.

More recently, the problem of formation of stable coalition structures
was considered in \cite{Yi97} in the presence of externalities and in
\cite{RV97} in the presence of binding agreements.  In both papers
two-stage games are analyzed. In the first stage coalitions form and
in the second stage the players engage in a non-cooperative game given
the emerged coalition structure.  In this context the question of
stability of the coalition structure is then analyzed.

Much research on stable coalition structures focussed on hedonic
games.  These are games in which the payoff of a player depends
exclusively on the members of the coalition he belongs to.  In other
words, a payoff of a player is a preference relation on the sets of
players that include him.  \cite{BJ02} considered four forms of
stability in such games: core, Nash, individual stability and
contractually individual stability.  Each alternative captures the
idea that no player, respectively, no group of playes has an incentive
to change the existing coalition structure.  The problem of existence
of (core, Nash, individually and contractually individually) stable
coalitions was considered in this and other
references, for example \cite{SBK01} and \cite{BZ03}.  A potentially
infinitely long coalition formation process in the context of hedonic games
was studied in \cite{BD05}.  This leads to another notion of
stability analogous to subgame perfect equilibrium.

Recently, \cite{BJ05} compared various notions of stability
and equilibria in network formation games. These are games in which the
players may be involved in a network relationship that, as a graph, may evolve.
Other interaction structures which players can form were considered in
\cite{Dem04}, in which formation of hierarchies was studied, and
\cite{MSPCP04} in which only bilateral agreements that follow a
specific protocol were allowed.

Finally, the computer science perspective is illustrated by \cite{CB04}
in which an approach to coalition formation 
based on Bayesian reinforcement was considered and tested empirically.

In this paper we propose to study the existence and formation of
stable coalition structures in the setting of coalitional games, by
proposing and studying a concept that bears some similarity with 
the Nash
equilibrium.  We consider partitions of the grand coalition and view a
partition stable if no group of players has a viable alternative to
staying within the partition.  The alternatives are provided as a set
of different coalition arrangements for the group of defecting players
and are compared using their social welfare.

The following example hopefully clarifies our approach.
We shall return to it in the last section.

\begin{example} \label{exa:chain}
  Consider a set of stores located in a number of cities.  Each
  store belongs to a chain.  Suppose that each chain has a
  contract with a transportation company to deliver goods to all the
  stores belonging to the chain.
  
  We can now envisage a situation in which a group of stores
  decides to leave this transportation arrangement and choose another
  one, for example the one in which the stores from the same
  city have a contract with one transportation company.

So from the viewpoint of the transportation logistics the stores
from the `defecting' group of stores are now partitioned not
according to the chains they belong to but according to the cities
they are located at.  Such an alternative transportation arrangement
is then a different, preferred, partition for the defecting group of
stores.  From our viewpoint the original transportation
arrangement was then unstable.  
\HB
\end{example}

These alternatives to the existing partition of the grand coalition
are formalized by means of a \emph{defection function} that assigns to
each partition a set of partitioned subsets of the grand coalition.
By considering different defection functions we obtain different
notions of stability.  Two most natural defection functions are the
one that allows formation of all partitions of all subsets and the one
that allows formation of all partitions of the grand coalition.

We characterize these notions of stability and in the second part of
the paper analyze the problem of whether and how so defined stable
partitions can be reached from any initial partition by means of
`local' transformations.

\section{Preliminary definitions}

We begin by introducing the basic concepts.
Let $N = \{1,2,\ldots ,n\}$ be a fixed set of players called the \textit{grand coalition} and let
$(v,N)$ be a coalitional TU-game (in short a game). That is,
$v$ is a function from the powerset of $N$ to ${\cal R}$.  
In what follows we assume that $v(\emptyset) = 0$.
We call the elements of $N$ \textit{players} and non-empty subsets of $N$ \textit{coalitions}.

A game $(v, N)$ is called 

\begin{itemize}

\item \textit{additive} if $v(A) + v(B) = v(A\cup B)$, 

\item \textit{superadditive} if $v(A) + v(B) \leq v(A\cup B)$

\item \textit{strictly superadditive} if
$v(A) + v(B) < v(A\cup B)$,
\end{itemize}
where in each case the condition holds for
every two disjoint coalitions $A$ and $B$ of $N$.

A \textit{collection} (in the grand coalition $N$) is any family $C :=
\{C_1,\ldots, C_l\}$ of mutually disjoint coalitions of
$N$, and $l$ is called its \textit{size}. If additionally
$\bigcup_{j=1}^l C_j = N$, the collection $C$ is called a
\textit{partition} of $N$.

\begin{figure}[htbp]
\centering
\input{coa.pstex_t}
\caption{A collection $C$ in the frame of a partition $P$}
\label{fig:coa}
\end{figure}
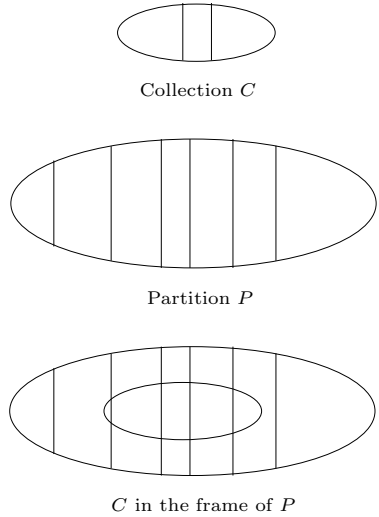

Given a collection $C := \{C_1,\ldots, C_l\}$ and a partition $P :=
\{P_1,\ldots, P_k\}$ we define
\[
C[P] := \{P_1 \cap (\cup_{j=1}^l C_j), \ldots,P_k \cap (\cup_{j=1}^l C_j)\} \setminus \{\emptyset\}
\]
and call $C[P]$ 
the \textit{collection} $C$ \textit{in the frame of} $P$.  
(By removing the empty set we ensure that $C[P]$ is a collection.)
To clarify this concept consider Figure \ref{fig:coa}. We
depict in it a collection $C$, a partition $P$ and $C$ in the frame of
$P$ (together with $P$). Here $C$ consists of three coalitions, while
$C$ in the frame of $P$ consists of five coalitions.

Intuitively, given a subset $S$ of $N$ and a partition $C:=
\{C_1,\ldots, C_l\}$ of $S$, the collection $C$ offers the players
from $S$ the `benefits' resulting from the partition of $S$ by $C$.
However, if a partition $P$ of $N$ is `in force', then the players
from $S$ enjoy instead the benefits resulting from the partition of
$S$ by $C[P]$, i.e., by $C$ in the frame of $P$.
Finally, note that if $C$ is a partition of $N$, then $C[P]$ is simply
$P$.

For a collection $C := \{C_1,\ldots, C_l\}$ we define now
\[
sw(C) := \sum_{j=1}^l v(C_j)
\]
and call $sw(C)$ the \textit{social welfare} of $C$.  
So for a partition $P := \{P_1,\ldots, P_k\}$
\[
sw(C[P]) = \sum_{i=1}^k v(P_i\cap (\cup_{j=1}^l C_j)).
\]
We call $sw(C[P])$
the $P$-\textit{modified social welfare} of the collection $C$.
That is, the $P$-modified social welfare of $C$ is the social
welfare of $C$ in the frame of $P$.

Given a partition $P := \C{P_1, \ldots, P_k}$ we call a coalition $T$ of
$N$ $P$-\textit{compatible} if for some $i \in \{1, \ldots, k\}$ we have $T
\sse P_i$ and $P$-\textit{incompatible} otherwise.
Further, we call a partition $Q := \{Q_1,\ldots, Q_l\}$
$P$-\textit{homogeneous} if for each $j \in \{1, \ldots, l\}$ some $i \in
\{1, \ldots, k\}$ exists such that either $Q_j \subseteq P_i$ or $P_i
\subseteq Q_j$.  Equivalently, a partition $Q$ is $P$-homogeneous if each $Q_j$ is
either $P$-compatible or equals $\cup_{i\in T} P_i$ for some
$T\subseteq \{1,\ldots , k\}$.  So any $P$-\textit{homogeneous}
partition arises from $P$ by allowing each coalition either to split
into smaller coalitions or to merge with other coalitions.

The crucial notion in our considerations is that of a \textit{defection
  function}. It is a function $\mathbb{D}$ which associates with each
partition $P$ a family of collections in $N$.  Intuitively we
interpret $\mathbb{D}$ as follows. For a partition $P$ the family
$\mathbb{D}(P)$ consists of all the collections $C := \{C_1,\ldots,
C_l\}$ whose players can leave the partition $P$ by forming a new,
separate, group of players $\cup_{j=1}^l C_j$ divided according to the
collection $C$.

Let $\mathbb{D}$ be a defection function. A partition $P$ of the grand coalition
$N$ is said to be $\mathbb{D}$-\textit{stable} if
\begin{equation}
\label{EQ_1}
sw(C[P]) \geq sw(C)
\end{equation}
for all collections $C \in \mathbb{D}(P)$.

If $C$ is a partition, then, as already noted $C[P] = P$, so
the above inequality reduces to
the comparison of the social welfare of $P$ and $C$:
\begin{equation}
  \label{eq:2}
sw(P) \geq sw(C).
\end{equation}

This definition has the following natural interpretation.  A
partition $P$ is $\mathbb{D}$-stable if no group of players is
interested in leaving $P$ when the players who wish to leave
can only form
collections allowed by the defection function $\mathbb{D}(P)$.  This
is a consequence of the fact that the `departing' players cannot
improve upon their social welfare in comparison with their $P$-modified social welfare.

In what follows we shall consider three natural types of defection functions
$\mathbb{D}$. Each will yield a different notion of a
$\mathbb{D}$-stable partition.


\section{$\mathbb{D}_c$-stability}

We begin with the defection function $\mathbb{D}_c$, where for each
partition $P$, $\mathbb{D}_c(P)$ is the family of all collections in
$N$.  So this defection function $\mathbb{D}_c$ allows any group of
players to leave $P$ and create an arbitrary collection in $N$.
So by definition a partition $P$ is
$\mathbb{D}_c$-stable if and only if (\ref{EQ_1}) holds
for every collection $C$ in $N$.

The following result shows that the $\mathbb{D}_c$-stability
condition can be considerably simplified.

\begin{theorem}
\label{T_1}
A partition $P := \{P_1,\ldots, P_k\}$ of $N$ is $\mathbb{D}_c$-stable if
and only if the following two conditions are satisfied:

\begin{enumerate}\smallromani
\item for each $i \in \{1, \ldots, k\}$ and each pair of disjoint coalitions
$A$ and $B$ such that $A \cup B \subseteq P_i$
\begin{equation}
\label{EQ_2}
v(A \cup B) \geq v(A) + v(B),
\end{equation}

\item for each $P$-incompatible coalition $T\subseteq N$
\begin{equation}
\label{EQ_3}
\sum_{i=1}^k v( P_i\cap  T) \geq v(T).
\end{equation}
\end{enumerate}
\end{theorem}

\noindent
\Proof

\noindent
($\Rightarrow$)
Suppose that $A \cup B \subseteq P_i$ for some $i$
and a pair $A$ and $B$ of disjoint coalitions. 
By taking the collection $C := \{A,B\}$ we get that (\ref{EQ_2}) 
is an immediate consequence of (\ref{EQ_1}) since by
assumption $v(\emptyset) = 0$.

Now suppose that $T\subseteq N$ is a $P$-incompatible coalition.
Then by taking the singleton collection $C := \{T\}$
we get that (\ref{EQ_3}) is an immediate consequence of (\ref{EQ_1}).
\medskip

\noindent
($\Leftarrow$) First note that (\ref{EQ_2}) implies that for each 
$i \in \{1, \ldots, k\}$ and for each collection $C := \{C_1,\ldots, C_l\}$ with 
$l > 1$ and $\cup_{j=1}^l C_j \subseteq P_i$
\begin{equation}
\label{EQ_2alt}
v(\cup_{j=1}^l C_j) \geq \sum_{j=1}^l v(C_j).
\end{equation}
Let now $C := \{C_1,\ldots, C_l\}$ be an arbitrary collection in $N$.
Define
$D^i := \{T \mid T\in C$ and $T\subseteq P_i\}$ for $i \in \{1, \ldots, k\}$.
So $D^i$ is the set of those elements of $C$ that are subsets of $P_i$.

Further, let $E := C\setminus \cup_{i=1}^k D^i$. So $C \setminus E =
\cup_{i=1}^k D^i$ and hence
$$
\sum_{i=1}^k \sum_{T\in D^i} v(T)  =  \sum_{T\in C\setminus E} v(T).
$$
Finally, let $E^i := \{T\cap P_i \mid T\in E\}$
for $i \in \{1, \ldots, k\}$.
By definition for every $i \in \{1, \ldots, k\}$ and  $T\in D^i \cup E^i$
we have $T\subseteq P_i$. Hence 
by (\ref{EQ_2alt}) for $i \in \{1, \ldots, k\}$
\[
v(\cup_{T\in D^i \cup E^i} \: T) \geq \sum_{T\in D^i\cup E^i} v(T).
\]

Moreover, for $i \in \{1, \ldots, k\}$ we have
$$
P_i \cap(\cup_{j=1}^l C_j) = \cup_{T\in D^i \cup E^i} \: T,
$$
so
$$
\sum_{i=1}^k v(P_i\cap (\cup_{j=1}^l C_j)) \geq \sum_{i=1}^k \sum_{T\in D^i\cup E^i} v(T).
$$

Further, since $\cup_{i=1}^k D^i = C \setminus E$, using (\ref{EQ_3})
we have the following chain of (in)equalities:
$$
\sum_{i=1}^k \sum_{T\in D^i\cup E^i} v(T) = \sum_{T\in C\setminus E} v(T) + \sum_{i=1}^k\sum_{T\in E^i}  v(T)
$$
$$
    =  \sum_{T\in C\setminus E} v(T) + \sum_{i=1}^k \sum_{T\in E} v(P_i \cap T) \geq
        \sum_{T\in C\setminus E} v(T) + \sum_{T\in E} v(T)  =  \sum_{j=1}^l v(C_j)\,.
$$
So we have shown that conditions (\ref{EQ_2}) and (\ref{EQ_3}) imply (\ref{EQ_1}).
\hfill $\Box$
\medskip

The following observation characterizes the games in which each
partition is $\mathbb{D}_c$-stable.

\begin{theorem} \label{thm:additive}
A game $(v, N)$ is additive if and only if
each partition is $\mathbb{D}_c$-stable.
\end{theorem}
\Proof
($\Rightarrow$)
By Theorem \ref{T_1}.
\medskip

\noindent
($\Leftarrow$)
Take two disjoint coalitions $A$ and $B$ of $N$. First consider a partition of $N$
which includes $A \cup B$ as an element. By item $(i)$ of Theorem \ref{T_1}
we then get $v(A\cup B) \geq v(A) + v(B)$.
Next, consider a partition $P$ of $N$ which includes $A$ and $B$ as elements.
Then $A \cup B$ is $P$-incompatible so by item $(ii)$ of Theorem \ref{T_1}
we get then $v(A) + v(B) \geq v(A \cup B)$.
So $(v,N)$ is additive.
\HB
\VV

In turn, the following observation shows that only few partitions can be
$\mathbb{D}_c$-stable.

\begin{note} \label{not:max}
  If $P$ is a $\mathbb{D}_c$-stable partition, then
  \begin{equation}
    \label{eq:max}
sw(P) = \max_Q sw(Q),    
  \end{equation}
where the maximum is taken over all partitions $Q$ in $N$. 
\end{note}
\Proof
For any partition $C$ of $N$
(\ref{EQ_1}) reduces to (\ref{eq:2}).
\HB
\VV

In fact, in general $\mathbb{D}_c$-stable partitions do not need to exist.

\begin{example} \label{exa:a}
Consider the game $(v, N)$ with $N=\{1,2,3\}$ and $v$ defined by:

\[
v(S) := \left\{ 
\begin{tabular}{ll}
2 &  \mbox{if $|S| = 1$} \\
5 &  \mbox{if $|S| = 2$} \\
6 &  \mbox{if $|S| = 3$}
\end{tabular}
\right . 
\]

We now show that no partition of $N$ is
$\mathbb{D}_c$-stable.  By Note \ref{not:max} it suffices to check
that no partition that maximizes the social welfare in the set of all
partitions is $\mathbb{D}_c$-stable.

First note that the social welfare of a partition $P$ is maximized
when $|P|=2$ ---in that case $sw(P) = 7$.  Then $P= \{P_1,P_2\}$ with
$|P_1|=2$ and $|P_2|=1$. Suppose now $P$ is $\mathbb{D}_c$-stable.
We can assume that $P_1=\{1,2\}$ and
$P_2=\{3\}$. (The other subcases are symmetric since $v$ is
symmetric.) Now putting $T :=\{2,3\}$ in condition (\ref{EQ_3}) 
of Theorem \ref{T_1} we get $2+2\geq
5$, which is a contradiction.
\HB
\end{example}

For specific types of games the $\mathbb{D}_c$-stable partitions do
exist, as the following two direct corollaries to Theorem \ref{T_1}
and Note \ref{not:max} show.

\begin{corollary}
  The one element partition $P=\{N\}$ is $\mathbb{D}_c$-stable if and
  only if the game $(v, N)$ is superadditive. Moreover, when $(v, N)$
  is strictly superadditive, then $P$ is a unique
  $\mathbb{D}_c$-stable partition.
\end{corollary}

\begin{corollary}
  The partition $P=\{\{1\}, \{2\}, \ldots , \{n\}\}$ is
  $\mathbb{D}_c$-stable if and only if the inequality $\sum_{i\in
    T}v(\{i\})\geq v(T)$ holds for all $T\subseteq N$.
Moreover, when all inequalities are strict, then
$P$ is a unique $\mathbb{D}_c$-stable partition.
\end{corollary}


\section{$\mathbb{D}_p$-stability}

We now consider a weaker version of stability for which stable
partitions are guaranteed to exist.  To this end we consider the
defection function $\mathbb{D}_p$, where for each partition $P$,
$\mathbb{D}_p(P)$ is the family of all partitions of $N$.  So the
defection function $\mathbb{D}_p$ allows a group of players to leave
$P$ only as the group of all players. They can form then an arbitrary
partition of $N$.

We have the following simple result.

\begin{theorem}
\label{thm:Dp}
A partition $P$ is $\mathbb{D}_p$-stable if and only if
(\ref{eq:max}) holds, 
where the maximum is taken over all partitions $Q$ in $N$. 
\end{theorem}

\Proof
For any partition $C$ of $N$
(\ref{EQ_1}) reduces to (\ref{eq:2}).
\HB
\VV

So the $\mathbb{D}_p$-stable partitions are exactly those that
maximize the social welfare in the set of all partitions. Consequently
the set of $\mathbb{D}_p$-stable partitions is non-empty.  However, in
specific situations the notion of a $\mathbb{D}_p$-stable partition
may be inadequate.  

As an example suppose there are $k$ locations and
we wish to associate `optimally' each player with a location, where
optimality means that the resulting social welfare is maximized.
It may easily happen that in the considered game
all $\mathbb{D}_p$-stable partitions have more than $k$ coalitions,
so an optimal partition in the above sense
cannot be described as a $\mathbb{D}_p$-stable partition.
To cope with such situations we modify this notion as follows.

Given $k \in \{1, \ldots, n\}$ let $\mathbb{D}_p^k$ be the defection
function such that for each partition $P$, $\mathbb{D}_p^k(P)$ is the
family of all partitions of size at most $k$. Then the
following result holds.

\begin{theorem}
\label{T_2}
For each $k \in \{1, \ldots, n\}$ there exists a partition $P$ of size
at most $k$ which is $\mathbb{D}_p^k$-stable.
\end{theorem}
\Proof Let $P$ 
be any partition of size at most $k$ and satisfying the equality
\[
sw(P) =  \max_{Q^k} sw(Q^{k}),
\]
where the maximum is taken over all partitions $Q^k$ of size at most $k$.
Then $P$ is the desired partition.
\hfill $\Box$


\section{$\mathbb{D}_{hp}$-stability}

Finally, we focus on the defection function $\mathbb{D}_{hp}$, where
for each partition $P$, $\mathbb{D}_{hp}(P)$ is the family of all
$P$-homogeneous partitions in $N$. So the defection function
$\mathbb{D}_{hp}$ allows the players to leave the partition $P$ only
by means of (possibly multiple) merges or splittings.  The existence
of a $\mathbb{D}_{hp}$-stable partition is then guaranteed by Theorem
\ref{thm:Dp} since every $\mathbb{D}_{p}$-stable partition is also 
$\mathbb{D}_{hp}$-stable. 
Moreover, the following obvious analogue of Theorem \ref{thm:Dp}
holds.

\begin{theorem}
\label{thm:Dp1}
A partition $P$ is $\mathbb{D}_{hp}$-stable if and only if
(\ref{eq:max}) holds, 
where the maximum is taken over all $P$-homogeneous
partitions $Q$ in $N$. 
\end{theorem}

Also, this notion of stability admits the following characterization.

\begin{theorem}
\label{T_1a}
A partition $P := \{P_1,\ldots, P_k\}$ of $N$ is $\mathbb{D}_{hp}$-stable if
and only if the following two conditions are satisfied:

\begin{enumerate}\smallromani
\item 
for each $i \in \{1, \ldots, k\}$ and for each partition $\{C_1,\ldots, C_l\}$ of
the coalition $P_i$
\begin{equation}
\label{EQ_2a}
v(P_i) \geq \sum_{j=1}^l v(C_j),
\end{equation}

\item for each $T\subseteq \{1, \ldots, k\}$
\begin{equation}
\label{EQ_3a}
\sum_{i\in T} v( P_i) \geq v(\cup_{i\in T} P_i).
\end{equation}
\end{enumerate}
\end{theorem}

\noindent
\Proof \\
($\Rightarrow$)
Let $C := \{C_1,\ldots, C_l\}$ be an arbitrary partition of some coalition $P_i$ in $P$.
Hence $\cup_{j=1}^l C_j = P_i$ and (\ref{EQ_2a}) is an immediate consequence of
(\ref{EQ_1})  applied
to the $P$-homogenous partition 
$\{C_1,\ldots, C_l, P_1, \ldots, P_{i-1}, P_{i+1}, \ldots, P_n\}$.

Next, consider some $T\subseteq N$. Then by applying (\ref{EQ_1}) to
the partition $C := \{\cup_{i\in T} P_i \} \cup \{P_i \mid i \not\in T\}$ we get 
the inequality (\ref{EQ_3a}).
\medskip

\noindent
($\Leftarrow$)
Let $P := \{P_1,\ldots, P_k\}$ be any partition of $N$ for which
conditions (i) and (ii) hold, and let $C := \{C_1,\ldots, C_l\}$ be an
arbitrary collection in $\mathbb{D}_{hp}(P)$.  By definition $C$ is a
$P$-homogeneous partition. Therefore the coalitions of $P$ can be
divided into disjoint groups $\{G_1,\ldots, G_r\}$ of collections as
follows: $P_i$ and $P_j$ belong to the same group if and only if
$P_i\cup P_j\subseteq C_s$ for some $s$.

Let us define $L=\{t \mid |G_t|=1\}$ and $M=\{t \mid |G_t|> 1\}$. By the $P$-homogeneity of
partition $C$, for each $t\in L$ there is a set, say $H_t$, such that
$P_t = \cup_{j\in H_t}C_j$. Similarly, for each $t\in M$ there is a coalition of $C$,
say $C_{q(t)}$, such that $C_{q(t)} = \cup_{P_i\in G_t}P_i$.
Hence, using (\ref{EQ_2a}) and (\ref{EQ_3a}), we get
the following chain of (in)equalities:

\[
\sum_{i=1}^k v(P_i\cap (\cup_{j=1}^l C_j)) = \sum_{t=1}^k v(P_t) =
\sum_{t\in L} v(\cup_{j\in H_t}C_j)  + \sum_{t\in M} \sum_{P_i\in G_t} v(P_i)
\]
\[
 \geq \sum_{t\in L} \sum_{j\in H_t} v(C_j)  + \sum_{t\in M} v(\cup_{P_i\in G_t} P_i) =
   \sum_{t\in L} \sum_{j\in H_t} v(C_j)  + \sum_{t\in M} v(C_{q(t)}) = \sum_{j=1}^l v(C_j)\,.
\]
So we have shown that conditions (\ref{EQ_2a}) and (\ref{EQ_3a}) imply (\ref{EQ_1}).
\HB
\VV

To summarize the relationship between the considered notions
of stable partition, given a defection function $\mathbb{D}$
denote by ${\cal ST}(\mathbb{D})$ the set of $\mathbb{D}$-stable partitions.
We have then the following obvious inclusions:
\[
{\cal ST}(\mathbb{D}_c) \subseteq  {\cal ST}(\mathbb{D}_p) \subseteq  {\cal ST}(\mathbb{D}_{hp}).
\]
With the possible exception of ${\cal ST}(\mathbb{D}_c)$ the considered sets of stable
partitions are always non-empty.

\section{Finding stable partitions}
\label{sec:strict}

Next we consider the problem of finding  stable partitions studied in the previous
sections.  To this end we introduce two rules that allow us to modify a partition of
$N$.\\

\noindent {\bf merge}
$$
\{T_1,\ldots, T_k\}\cup P \rightarrow \{\cup_{j=1}^k T_j \} \cup P,
$$

where $\sum_{j=1}^k v(T_j) < v(\cup_{j=1}^k T_j)$,
\\

\noindent {\bf split}
$$
\{\cup_{j=1}^k T_j \} \cup P \rightarrow  \{T_1,\ldots, T_k\}\cup P,
$$

where $\{T_1,\ldots, T_k\}$ is a collection such that
 $v(\cup_{j=1}^k T_j) < \sum_{j=1}^k v(T_j)$,
\\

The following observation holds.

\begin{note}
\label{not:1}
Every iteration of the merge and split rules terminates.
\end{note}
\Proof
This is an immediate consequence of the fact that each rule application increases
the social welfare.
\HB
\VV

We now proceed with the characterizations of the introduced notions of
stable partitions using the above rules. The cases of $\mathbb{D}_c$-stable
partitions and $\mathbb{D}_p$-stable partitions are not so
straightforward, so we begin with the notion of $\mathbb{D}_{hp}$-stability.
The following result holds.

\begin{theorem} \label{T_4} 

A partition is $\mathbb{D}_{hp}$-stable if and only if it is the
outcome of iterating the merge and split rules.
\end{theorem}

\noindent
\Proof 
It is an immediate consequence of Theorem \ref{T_1a}.
\HB
\VV

So to find a $\mathbb{D}_{hp}$-stable partition it suffices to iterate
the merge and split rules starting from any initial partition until
one reaches a partition closed under the applications of these rules.

In general the outcome of various iterations of the merge and split
rules does not need to be unique.  Moreover, some of these outcomes do not have a
maximal social welfare.





\begin{example} \label{exa:1}
Consider the following game $(v, N)$. Let $N = \C{1,2,3,4}$ and let
$v$ be defined as follows:

\[
v(S) := \left\{ 
\begin{tabular}{ll}
1 &  \mbox{if $S = \{1, 2\}$} \\
2 &  \mbox{if $S = \{1, 3\}$} \\
0 &  \mbox{otherwise}
\end{tabular}
\right . 
\]

Consider now the partition $\C{\C{1}, \C{2}, \C{3}, \C{4}}$ of $N$.
By the merge rule we can transform it to $\C{\C{1,2}, \C{3}, \C{4}}$
or to $\C{\C{1,3}, \C{2}, \C{4}}$.  The social welfare of these
partitions is, respectively, 1 and 2.  In each case we reached a
partition to which neither merge nor split rule can be applied.
\HB
\end{example}

We now proceed with an analysis of $\mathbb{D}_c$-stable partitions
using the merge and split rules. First note the following observation.

\begin{note} \label{not:stable}
Every $\mathbb{D}_c$-stable partition is closed under the applications
of the merge and split rules.
\end{note}
\Proof
This is an immediate consequence of Note \ref{not:max}.
\HB
\VV

Unfortunately it is not possible to characterize $\mathbb{D}_c$-stable
partitions using the merge and split rules.  First, not every partition closed
under the applications of the merge and split rules is
$\mathbb{D}_c$-stable.  Indeed, take the game from Example \ref{exa:a}
and consider a partition with the maximal social welfare.  This
partition is closed under the applications of the merge and split
rules but we saw already that in this game no
$\mathbb{D}_c$-stable partition exists.

Second, there are games in which $\mathbb{D}_c$-stable partitions exist,
but some iterations of the merge and split rules may miss them.

\begin{example} \label{exa:miss}
Let $N = \C{1,2,3,4}$ and define
$v$ as follows:

\[
v(S) := \left\{ 
\begin{tabular}{ll}
3 &  \mbox{if $S = \{1,2\}$} \\
$|S|$ &  \mbox{otherwise}
\end{tabular}
\right . 
\]

It is straightforward to show using Theorem \ref{T_1} that 
$\{\{1,2\}, \{3,4\}\}$ is a $\mathbb{D}_c$-stable partition.  Its social welfare is
5.  Now, the partition $\{\{1,3\}, \{2,4\}\}$ is 
closed under the applications of the merge and split rules.
But its social welfare is 4 so by Note \ref{not:max} 
it is not $\mathbb{D}_c$-stable.
\HB
\end{example}

Note that in the above example $\{\{1,2\}, \{3,4\}\}$ is also a unique
$\mathbb{D}_p$-stable partition (since it is the only partition with
the social welfare 5), so some iterations of the merge and split
rules may also miss the $\mathbb{D}_p$-stable partitions.

Third, there are games in which all iterations of the merge and split
rules have a unique outcome (which happens to be the unique
partition that maximizes the social welfare in the set of all
partitions), yet no $\mathbb{D}_c$-stable partition exists.

\begin{example} \label{exa:2}
Consider the following game. Let $N = \C{1,2,3,4}$ and let
$v$ be defined as follows:

\[
v(S) := \left\{ 
\begin{tabular}{ll}
6 &  \mbox{if $S = \C{1,2,3,4}$} \\
4 &  \mbox{if $S = \C{1,2}$ or $S = \C{3,4}$} \\
3 &  \mbox{if $S = \C{1,3}$} \\
$|S|$ &  \mbox{otherwise}
\end{tabular}
\right . 
\]

Clearly neither merge nor split rule can be applied to
the partition $\C{\C{1,2}, \C{3,4}}$.
We now show that $\C{\C{1,2}, \C{3,4}}$ 
is a unique outcome of the 
iterations of the merge and split rules.
Take a partition $\C{T_1, \ldots, T_k}$ different from 
$\C{\C{1,2}, \C{3,4}}$. 

If for some $i$ we have 
$T_i = \C{1,2}$, then $\C{T_1, \ldots, T_k} = \C{\C{1,2}, \C{3}, \C{4}}$ and
consequently
$\C{\C{1,2}, \C{3}, \C{4}} \myra \C{\C{1,2}, \C{3,4}}$
by the merge rule.
If for some $i$ we have $T_i = \C{3,4}$, then the argument is symmetric.
If for some $i$ we have 
$T_i = \C{1,3}$, then for $j \neq i$ we have $v(T_j) = |T_j|$.
Consequently, by the merge and split rule
\[
\C{T_1, \ldots, T_k} \myra \C{\C{1,2,3,4}} \myra \C{\C{1,2}, \C{3,4}}.
\]

If $\C{T_1, \ldots, T_k} = \C{\{1,2,3,4\}}$, then by the split rule
\[
\C{\C{1,2,3,4}} \myra \C{\C{1,2}, \C{3,4}}.
\]

If no $T_i$ equals $\C{1,2}, \C{3,4}, \C{1,3}$ or $\C{1,2,3,4}$, then for all $i$ we
have $v(T_i) = |T_i|$ and consequently by the merge rule and by the split rule
\[
\C{T_1, \ldots, T_k} \myra \C{\C{1,2,3,4}} \myra \C{\C{1,2}, \C{3,4}}.
\]

So in this game all iterations of the merge and split rules have a
unique outcome, $P := \C{\C{1,2}, \C{3,4}}$. 
By Note \ref{not:stable} $P$
is the only possible $\mathbb{D}_c$-stable partition.  However, for
the $P$-incompatible set $\C{1,3}$ we have
\[
v(\C{1,3}) > v(\C{1}) + v(\C{3}) = v(\C{1,3} \cap \C{1,2}) + v(\C{1,3} \cap \C{3,4}).
\]
So by Theorem \ref{T_1} $P$ is not $\mathbb{D}_c$-stable and
consequently this game has no $\mathbb{D}_c$-stable partitions.  
\HB
\end{example}

A natural question then arises
whether some other simple rules exist using which we could
characterize the $\mathbb{D}_{c}$-stable and $\mathbb{D}_{p}$-stable
partitions.

This is very unlikely. To clarify the matters 
let us return to Example \ref{exa:miss}.
We noted there that the partition $\{\{1,3\}, \{2,4\}\}$ (with social
welfare 4) is not $\mathbb{D}_c$-stable but is closed the applications
of the merge and split rules.  

So to transform $\{\{1,3\}, \{2,4\}\}$
to $\{\{1,2\}, \{3,4\}\}$, which is a unique $\mathbb{D}_c$-stable and
$\mathbb{D}_p$-stable partition (and with social welfare 5), we need
more powerful rules.  As we limit ourselves to rules that lead to a
strict increase of the social welfare, the only solution is to
transform $\{\{1,3\}, \{2,4\}\}$ to $\{\{1,2\}, \{3,4\}\}$ directly,
in one rule application.
In particular, the following natural rule does not suffice:
\\

\noindent {\bf transfer}
$$
\{T_1, T_2\} \cup P \rightarrow  \{T_1\setminus U, \,T_2\cup U\}\cup P,
$$

where $U \subset T_1$ and $v(T_1)+v(T_2) < v(T_1\setminus U) + v(T_2\cup U)$.
\\

What we need is a rule that leads to a `bidirectional transfer', for example
\\

\noindent {\bf exchange}
$$
\{T_1, T_2\} \cup P \rightarrow  \{T_1 \setminus U_1 \cup U_2, \,T_2 \setminus U_2 \cup U_1\}\cup P,
$$

where $U_1 \subset T_1$,  $U_2 \subset T_2$  
and $v(T_1)+v(T_2) < v(T_1 \setminus U_1 \cup U_2) + v(T_2  \setminus U_2 \cup U_1)$.
\\

This rule suffices here. But it is easy to construct an example where this
rule does not suffice either. We just need to generalize appropriately Example
\ref{exa:miss}.

Let $N := \{1, 2, \LL, 2 n$, where $n > 1$, and define 
$v$ as follows:

\[
v(S) := \left\{ 
\begin{tabular}{ll}
n+1 &  \mbox{if $S = \{1,3, \LL, 2  n - 1\}$} \\
$|S|$ &  \mbox{otherwise}
\end{tabular}
\right . 
\]

Using Theorems \ref{T_1} and \ref{thm:Dp}
it is straightforward to check that
$\{\{1,3, \LL, 2 n - 1\}, \{2,4, \LL, 2 n\}\}$ is the unique
$\mathbb{D}_c$-stable and unique $\mathbb{D}_p$-stable partition. Its social
welfare is $2 n+1$. Now the partition $\{P_1, \LL, P_n\}$, where $P_i
:= \{2 i - 1, 2 i\}$ for $i \in \{1, \LL, n\}$ has the social welfare
$2 n$.  So to transform it to $\{\{1,3, \LL, 2 n - 1\}, \{2,4, \LL, 2 n\}\}$
we need a rule that can achieve it in one application. Of
course such a rule exists: it suffices to group the odd numbers into
one set and the even numbers into another. But it should be clear that
any generic way of formulating this operation leads to a pretty
complex rule.

Continuing this line it is easy to envisage a series of increasingly
more complex examples which suggest that in the end the only rule
using which we can characterize the $\mathbb{D}_p$-stable partitions
seems to be the one that allows us to transform an arbitrary partition
into another one when the social welfare increases.  This defeats our
purpose of finding a characterization by means of simple rules.
The fact that the $\mathbb{D}_c$-stable partitions do not need to exist
makes the task of characterizing them by means of simple rules even more unlikely.

\section{Strictly stable partitions}
\label{sec:strictly}

As a way out of this dilemma we consider a more refined notion of a stable
partition.  Take a partition $P$ of $N$.
First note that for a collection $C$ that consists of some coalitions of $P$, i.e.,  
for $C \sse P$ we have $C[P] = C$ and consequently
(\ref{EQ_1}) then holds. 

Given a defection function  $\mathbb{D}$ we now say that
a partition $P$ of $N$ is \textit{strictly} $\mathbb{D}$-\textit{stable} if
\[
sw(C[P]) > sw(C)
\]
for all collections $C \in \mathbb{D}(P)$ that are not subsets of $P$.

We now analyze strictly $\mathbb{D}_c$-stable partitions.
The following analogue of Theorem \ref{T_1} holds.

\begin{theorem} 
\label{T_strict}
A partition $P := \{P_1,\ldots, P_k\}$ of $N$ is strictly $\mathbb{D}_c$-stable if
and only if the following two conditions are satisfied:
\begin{itemize}

\item for each $i \in \{1, \ldots, k\}$ and each pair of disjoint coalitions
$A$ and $B$ such that $A \cup B \subseteq P_i$
\begin{equation}
\label{EQ_a}
v(A \cup B) > v(A) + v(B),
\end{equation}

\item for each $P$-incompatible coalition $T\subseteq N$
\begin{equation}
\label{EQ_b}
\sum_{i=1}^k v( P_i \cap T) > v(T).
\end{equation}
\end{itemize}
\end{theorem}

Note that (\ref{EQ_a}) and (\ref{EQ_b}) are simply the sharp
counterparts of the inequalities (\ref{EQ_2}) and (\ref{EQ_3}).
\II

\noindent
\Proof
The proof is a direct modification of the proof of 
Theorem \ref{T_1}.
\HB
\VV

Next, we establish the following lemma.

\begin{lemma} \label{lem:unique}
Suppose that a strictly $\mathbb{D}_c$-stable partition $P$ exists. Let 
$P'$ be a partition which is
closed under the applications of the merge and split rules. Then
$P' = P$.
\end{lemma}

\Proof Suppose $P = \C{P_1, \ldots, P_k}$ and $P' = \C{T_1, \ldots, T_m}$.
Assume by contradiction that $\C{P_1, \ldots, P_k} \neq \C{T_1, \ldots,
  T_m}$.  Then $\te i_0 \in \{1, \ldots, k\} \: \fa j \in \{1, \ldots, m\}
\: P_{i_0} \neq T_j$.  Let $T_{j_1}, \ldots, T_{j_l}$ be the minimum
cover of $P_{i_0}$.  \II

\NI
\emph{Case 1}. $P_{i_0} = \cup_{h=1}^{l} T_{j_h}$.

Then $l > 1$ and $\C{T_{j_1}, \ldots, T_{j_l}}$ is a partition of
$P_{i_0}$.  But $P$ is strictly $\mathbb{D}_c$-stable, so by 
Theorem \ref{T_strict} and (\ref{EQ_a})
$\sum_{h=1}^{l} v(T_{j_h}) < v(\cup_{h=1}^{l} T_{j_h})$. 
Consequently, the merge rule can be applied to $\C{T_1,
\ldots, T_m}$, which is a contradiction.  
\II

\NI
\emph{Case 2}. $P_{i_0}$ is a proper subset of $\cup_{h=1}^{l} T_{j_h}$.

Then for some $j_h$ the set $P_{i_0} \cap T_{j_h}$ is a proper
non-empty subset of $T_{j_h}$.  So $T_{j_h}$ is a $P$-incompatible
set.  But $P$ is strictly $\mathbb{D}_c$-stable, so by 
Theorem \ref{T_strict} and (\ref{EQ_b}) $v(T_{j_h})
< \sum_{i=1}^{k} v(P_i \cap T_{j_h})$.  Consequently, since $T_{j_h}
= \cup_{i=1}^{k} (P_i \cap T_{j_h})$, the split rule can be applied to
$\C{T_1, \ldots, T_m}$ which is a contradiction.  
\HB 
\VV

This allows us to draw the following conclusions.

\begin{theorem} \label{thm:sstable}
Suppose that $P$ is a 
strictly $\mathbb{D}_c$-stable partition. Then

\begin{enumerate} \smallromani

\item $P$ is the outcome of every iteration of the merge and split rules.

\item $P$ is a unique $\mathbb{D}_c$-stable partition.

\item $P$ is a unique $\mathbb{D}_p$-stable partition.

\item $P$ is a unique $\mathbb{D}_{hp}$-stable partition.

\end{enumerate}

\end{theorem}
\Proof 

\NI
$(i)$ By Note \ref{not:1}
every iteration of
the merge and split rules terminates,
so the claim  follows by Lemma \ref{lem:unique}. \\[2mm]
$(ii)$ Suppose that $P'$ is a $\mathbb{D}_c$-stable partition.
By Note \ref{not:stable} $P'$ is
closed under the applications of the merge and split rules, so 
by Lemma \ref{lem:unique}
$P' = P$. \\[2mm]
$(iii)$ By Theorem \ref{thm:Dp} 
a partition is $\mathbb{D}_p$-stable if and only if
it maximizes the social welfare (in the set of all partitions).
But each partition that maximizes the social welfare
is closed under the applications of the merge and split rules,
so the claim follows from $(i)$.
\\[2mm]
$(iv)$ By $(i)$ and Theorem \ref{T_4}.
\HB
\VV

Item $(i)$ shows that if a strictly $\mathbb{D}_c$-stable partition exists, then
we can reach it from any initial partition through an arbitrary
iteration of the merge and split rules.

Example \ref{exa:miss} shows that the concepts of unique and
strictly $\mathbb{D}_c$-stable partitions do not coincide.  Indeed, by
Note \ref{not:max} $\{\{1,2\}, \{3,4\}\}$ is there a unique
$\mathbb{D}_c$-stable partition.  However, Theorem \ref{T_strict}
implies that $\{\{1,2\}, \{3,4\}\}$ is not strictly
$\mathbb{D}_c$-stable.  This is in contrast to the
case of $\mathbb{D}_p$-stable partitions as the following
characterization result shows.

\begin{theorem} \label{thm:spstable}
A partition is strictly $\mathbb{D}_p$-stable if and only if
it is a unique $\mathbb{D}_p$-stable partition.
\end{theorem}

\noindent
\Proof \\
($\Rightarrow$)
Let $P$ be a strictly $\mathbb{D}_p$-stable partition.
By definition $sw(C[P]) > sw(C)$ for all partitions 
$C$ different from $P$, 
or equivalently $sw(P) > sw(P')$ for all partitions $P'$ different from $P$.
Let $P'$ be a $\mathbb{D}_p$-stable partition. By Theorem \ref{thm:Dp}
both $P$ and $P'$ maximize the social welfare in the set of all partitions.
So $sw(P) = sw(P')$ and consequently $P$ and $P'$ coincide. 
\medskip

\noindent
($\Leftarrow$)
Suppose $P$ is a unique $\mathbb{D}_p$-stable partition.
Let $P'$ be a partition different from $P$.
By Theorem \ref{thm:Dp} and uniqueness of $P$
$sw(P) > sw(P')$. So
$P$ is a strictly $\mathbb{D}_p$-stable partition.
\HB
\VV

The full analogue of Theorem \ref{thm:sstable} does not hold.
Indeed, consider the following modification of Example \ref{exa:miss}.

\begin{example} \label{exa:miss1}
Let $N = \C{1,2,3}$ and define
$v$ as follows:

\[
v(S) := \left\{ 
\begin{tabular}{ll}
3 &  \mbox{if $S = \{1,2\}$} \\
$|S|$ &  \mbox{otherwise}
\end{tabular}
\right . 
\]

By Theorems \ref{thm:Dp} and \ref{thm:spstable}
$\{\{1,2\}, \{3\}\}$ is a strictly
$\mathbb{D}_p$-stable partition.  Its social welfare is 4.  But the
partition $\{\{1,3\}, \{2\}\}$ is closed under the applications of the
merge and split rules and hence, by Theorem \ref{T_4}, is
$\mathbb{D}_{hp}$-stable.  So $\{\{1,2\}, \{3\}\}$ is not the outcome
of every iteration of the merge and split rules and is not a unique
$\mathbb{D}_{hp}$-stable partition.  
\HB
\end{example}

In turn, Example \ref{exa:2} shows that the existence of a strictly
$\mathbb{D}_p$-stable partition does not imply the existence of a
$\mathbb{D}_c$-stable partition.

Finally, the following result deals with the strictly
$\mathbb{D}_{hp}$-stable partitions.

\begin{theorem} \label{thm:hp}
\mbox{}\\[-6mm]
  \begin{enumerate}  \smallromani
  \item A partition is strictly $\mathbb{D}_{hp}$-stable if and only if
it is a unique $\mathbb{D}_{hp}$-stable partition.

\item 
A partition is strictly $\mathbb{D}_{hp}$-stable if and only if it is
strictly $\mathbb{D}_{p}$-stable.  
  \end{enumerate}
\end{theorem}
\Proof

\noindent
$(i)$ The proof is the same as that of Theorem \ref{thm:spstable}, relying on
Theorem \ref{thm:Dp1} instead of Theorem \ref{thm:Dp}. \\[2mm]
$(ii)$

\noindent
($\Rightarrow$)
Let $P$ be a strictly $\mathbb{D}_{hp}$-stable partition.
Take an arbitrary partition $P'$ different from $P$
and let $P''$ be an arbitrary closure
of $P'$ under the applications of the merge and split rules.
By Theorem \ref{T_4} $P''$ is
$\mathbb{D}_{hp}$-stable, so 
by the choice of $P$ either $P = P''$ or $sw(P) > sw(P'')$.

If $P = P''$, then, by the choice of $P'$, $P'$ is
different from $P''$, so $sw(P'') > sw(P')$ and consequently $sw(P) > sw(P')$.
In turn, if $sw(P) > sw(P'')$, then, since $sw(P'') \geq sw(P')$, we get
$sw(P) > sw(P')$, as well. \\[2mm]
($\Leftarrow$)
Directly by definition.
\HB

\section{Existence of stable partitions}

We saw already that in general a $\mathbb{D}_c$-stable partition does
not need to exist, so a strictly $\mathbb{D}_c$-stable
partition does not need to exist either.

Under what condition a strictly $\mathbb{D}_c$-stable partition does
exists?  First note that by definition if the game is (strictly)
superadditive, then $\C{N}$ is its (strictly) $\mathbb{D}_c$-stable
partition.  The following example introduces a natural class of
non-superadditive games in which a $\mathbb{D}_c$-stable
(respectively, a strictly $\mathbb{D}_c$-stable) partition exists.

\begin{example} 
  Consider a partition $P := \C{P_1, \ldots, P_k}$ of $N$. Let $(v, N)$ be a
  game which is non-negative (that is,
  $v(A) \geq 0$ for all coalitions $A$ of $N$), superadditive when limited to subsets of a coalition
  $P_i$ (for all $i \in \{1, \ldots, n\}$) and such that $v(A) = 0$ for
  all $P$-incompatible sets.  
  It is straightforward to check with the help of Theorem \ref{T_1}
  that $P$ is a (not necessarily unique ---see Theorem \ref{thm:additive})
  $\mathbb{D}_c$-stable partition. 
  
  But if we additionally stipulate that $(v, N)$ is positive (that is,
  $v(A) > 0$ for all coalitions $A$ of $N$) and strictly
  superadditive, in each case when limited to subsets of a coalition
  $P_i$ (for all $i \in \{1, \ldots, n\}$), then $P$ becomes a
  strictly $\mathbb{D}_c$-stable partition, and hence a unique
  partition with this property.

To see specific examples of such games choose a partition $\C{P_1, \ldots, P_k}$ of $N$
and fix $m \geq 1$. Let
\[
v(S) := \left\{ 
\begin{tabular}{ll}
$|S|^m$ &  \mbox{if $S \sse P_i$ for some $i$} \\
0 &  \mbox{otherwise}
\end{tabular}
\right . 
\]
Then when $m = 1$ we get an example of a game in the first category
and  when $m > 1$ we get an example of a game in the second category.
\HB
\end{example}

Next, Theorems \ref{thm:Dp} and \ref{thm:spstable} provide a simple
criterion for the existence of a strictly $\mathbb{D}_p$-stable
partition: such a partition exists if and only if exactly
one partition maximizes the social welfare in the set of all
partitions.
By virtue of Theorem \ref{thm:hp} the same existence criterion applies to
strictly $\mathbb{D}_{hp}$-stable partitions.

Finally, let us return to our initial Example \ref{exa:chain}.  
Assume that:

\begin{itemize}
\item within each city 
the transportation costs per store decrease as the number
of served stores increases (economy of scale),

\item the transportation costs per store are
always lower if the served stores are located in the same city,

\item the stores aim at minimizing the transportation costs
and that switching transportation companies incurs no costs.
\end{itemize}

To formally analyze this example 
assume that there are $n$ stores and that
$\{P_1, \LL, P_k\}$ is the partition of the stores per city.
Denote by $c(S)$ the total transportation costs to the
set of stores $S$ and let 
\[
v(S) := \sum_{i \in S} c(\{i\}) - c(S),
\]
i.e., $v(S)$ is the cost saving for coalition $S$.

Then the first assumption states that the function $c(S)/|S|$, when
limited to the sets of stores within one city, strictly decreases as
the size of the set of stores $S$ increases. This easily implies that
the game $(v, N)$ is strictly superadditive when limited to the set of
stores within one city (see \cite[pages 93-95]{Mou88}).\footnote{The
  argument is as follows. Let $S$ and $T$ be disjoint coalitions, both
  subsets of some $P_i$. For some $\alpha, \beta \in (0,1)$
\[
c(S \cup T)/|S \cup T| = \alpha \cdot c(S)/|S| = \beta \cdot c(T)/|T|.
\]
Let $\gamma := |S|/|S \cup T|$. Then $1 - \gamma = |T|/|S \cup T|$. 
Hence $\gamma \cdot c(S \cup T) = \alpha \cdot c(S)$ and
$(1 - \gamma) \cdot c(S \cup T) = \beta \cdot c(T)$.
Consequently 
\[
c(S \cup T) = \alpha \cdot c(S) + \beta \cdot c(T) < c(S) + c(T).
\]
So by definition of $v$ we get $v(S) + v(T) < v(S \cup T)$.}

In turn, the second assumption states that for each 
$P$-incompatible coalition $T\subseteq N$
(representing a set of stores from different cities)
we have for $i \in \{1, \LL, k\}$
\[
c(P_i \cap T)/|P_i \cap T| < c(T)/|T|,
\]
or equivalently $c(P_i \cap T) < c(T) \cdot |P_i \cap T|/|T|$.
This implies
$\sum_{i=1}^k c( P_i \cap T) < c(T)$, so
$\sum_{i=1}^k v( P_i \cap T) > v(T)$ by the definition of $v$.

So using Theorem \ref{T_strict}
we get that the partition $\{P_1, \LL, P_k\}$ of the stores per
city is strictly $\mathbb{D}_c$-stable. Using Theorem
\ref{thm:sstable}$(i)$ we now conclude that the initial
transportation arrangement, per chain, can be broken and will lead
through an arbitrary sequence of splits and merges to the alternative
transportation arrangement, per city.

\bibliographystyle{plain}

\bibliography{/ufs/apt/bib/e}
\end{document}

%% file: header.tex
\def\smallromani{\renewcommand{\theenumi}{\roman{enumi}}
\renewcommand{\labelenumi}{(\theenumi)}}

\newcommand{\Proof}{\NI
                    {\bf Proof.}\ }
\newtheorem{theorem}{Theorem}[section]
\newtheorem{defined}[theorem]{Definition}

\newtheorem{exa}[theorem]{Example}
\newenvironment{example}{\begin{exa} \rm}{\end{exa}}

\newtheorem{lemma}[theorem]{Lemma}
\newtheorem{corollary}[theorem]{Corollary}

\newtheorem{note}[theorem]{Note}

\newtheorem{exe}{Exercise}

\newtheorem{pro}{Problem}

\newcounter{symbol}
\newcommand{\indexsyma}[1]%
{\stepcounter{symbol}\index{zzz1 \thesymbol @\protect#1}}
\newcommand{\indexsymb}[1]%
{\stepcounter{symbol}\index{zzz2 \thesymbol @\protect#1}}
\newcommand{\indexsymc}[1]%
{\stepcounter{symbol}\index{zzz3 \thesymbol @\protect#1}}
\newcommand{\indexsymd}[1]%
{\stepcounter{symbol}\index{zzz4 \thesymbol @\protect#1}}
\newcommand{\indexsyme}[1]%
{\stepcounter{symbol}\index{zzz5 \thesymbol @\protect#1}}

%% file: ecmds.tex
\newcommand{\myra}{\mbox{$\:\rightarrow\:$}}

\newcommand{\sse}{\mbox{$\:\subseteq\:$}}

\newcommand{\fa}{\mbox{$\forall$}}
\newcommand{\te}{\mbox{$\exists$}}

\newcommand{\LL}{\mbox{$\ldots$}}

\newcommand{\C}[1]{\mbox{$\{{#1}\}$}}           

\newcommand{\NI}{\noindent}
\newcommand{\HB}{\hfill{$\Box$}}
\newcommand{\VV}{\vspace{5 mm}}

\newcommand{\II}{\vspace{2 mm}}

\newcommand{\szkew}[1]{\relax \setbox0=\hbox{\kern -24pt $\displaystyle#1$\kern 0pt }%
\box0}
{\catcode`\@=11 \global\let\ifjusthvtest@=\iffalse}

\newcounter{oldmycaption}



%% file: coa.pstex_t
\begin{picture}(0,0)%
\includegraphics{coa.pstex}%
\end{picture}%
\setlength{\unitlength}{2368sp}%
\begingroup\makeatletter\ifx\SetFigFont\undefined%
\gdef\SetFigFont#1#2#3#4#5{%
  \reset@font\fontsize{#1}{#2pt}%
  \fontfamily{#3}\fontseries{#4}\fontshape{#5}%
  \selectfont}%
\fi\endgroup%
\begin{picture}(3852,5386)(4406,-6635)
\put(5476,-6586){\makebox(0,0)[lb]{\smash{{\SetFigFont{7}{8.4}{\familydefault}{\mddefault}{\updefault}{\color[rgb]{0,0,0}$C$ in the frame of $P$}%
}}}}
\put(5776,-2236){\makebox(0,0)[lb]{\smash{{\SetFigFont{7}{8.4}{\familydefault}{\mddefault}{\updefault}{\color[rgb]{0,0,0}Collection $C$}%
}}}}
\put(5851,-4411){\makebox(0,0)[lb]{\smash{{\SetFigFont{7}{8.4}{\familydefault}{\mddefault}{\updefault}{\color[rgb]{0,0,0}Partition $P$}%
}}}}
\end{picture}%

%% file: ar06.bbl
\begin{thebibliography}{10}

\bibitem{AD74}
R.J. Aumann and J.H. Dr\`{e}ze.
\newblock Cooperative games with coalition structures.
\newblock {\em International Journal of Game Theory}, (3):217--237, 1974.

\bibitem{BD05}
F.~Bloch and E.~Diamantoudi.
\newblock Noncooperative formation in coalitions in hedonic games, 2005.
\newblock Working paper.

\bibitem{BJ05}
F.~Bloch and M.~Jackson.
\newblock Definitions of equilibrium in network formation, 2005.
\newblock Working paper.

\bibitem{BJ02}
A.~Bogomolnaia and M.~Jackson.
\newblock The stability of hedonic coalition structures.
\newblock {\em Games and Economic Behavior}, 38(2):201--230, 2002.

\bibitem{BZ03}
N.~Burani and W.S. Zwicker.
\newblock Coalition formation games with separable preferences.
\newblock {\em Mathematical Social Sciences}, 45(1):27--52, 2003.

\bibitem{CB04}
G.~Chaldiakis and C.~Boutilier.
\newblock Bayesian reinforcement learning for coalition formation under
uncertainty.
\newblock {\em AAMAS '04} Conference, 2004.

\bibitem{Dem04}
G.~Demange.
\newblock On group stability in hierarchies and networks.
 2004.
\newblock Working paper.

\bibitem{Gre94}
J.~Greenberg.
\newblock Coalition structures.
\newblock In R.J. Aumann and S.~Hart, editors, {\em Handbook of Game Theory
  with Economic Applications}, volume~2 of {\em Handbook of Game Theory with
  Economic Applications}, chapter~37, pages 1305--1337. Elsevier, 1994.

\bibitem{Mou88}
H.~Moulin.
\newblock Axioms of Cooperative Decision Making.
\newblock Cambridge University Press, 1988.

\bibitem{MSPCP04}
I.~Macho-Stadler, D.~P¨¦rez-Castrillo, and N.~Porteiro.
\newblock Sequential formation of coalitions through bilateral agreements,
  2005.
\newblock Working paper.

\bibitem{RV97}
D.~Ray and R.~Vohra.
\newblock Equilibrium binding agreements.
\newblock {\em Journal of Economic Theory}, (73):30--78, 1997.

\bibitem{SBK01}
T.~S\"{o}nmez, S.~Banerjee, and H.~Konishi.
\newblock Core in a simple coalition formation game.
\newblock {\em Social Choice and Welfare}, 18(1):135--153, 2001.

\bibitem{Yi97}
S.S. Yi.
\newblock Stable coalition structures with externalities.
\newblock {\em Games and Economic Behavior}, 20:201--237, 1997.

\end{thebibliography}
